 \documentclass[final,5p,times,twocolumn,authoryear]{elsarticle}

\usepackage{amssymb}
\usepackage{amsmath}
\usepackage{lipsum}
\begin{document}

\begin{frontmatter}


\author[1,2]{M.C. Gordillo\corref{cor1}\fnref{label2}}
\ead{cgorbar@upo.es}
\author[1] {J. Segovia}
\affiliation[1]{organization={Departamento de Sistemas Fisicos, Quimicos y Naturales, Universidad Pablo de Olavide},
          city={Sevilla},
          postcode={E-41013}, 
           country={Spain}}
           
\affiliation[2]{organization=Instituto Carlos I de Fisica Teorica y Computacional, Universidad de Granada. Granada. E-18071, Spain}

\title{Diffusion Monte Carlo calculation of compact $T_{cs0}$ and $T_{c\bar{s}0}$ 
tetraquarks}


%\affiliation[first]{organization={University of the Moon},%Department and Organization
%            city={Earth},
%\author[1] {M.C. Gordillo^*}
%\affiliation[1]{Departamento de Sistemas Fisicos, Quimicos %y Naturales, Universidad Pablo de Olavide, E-41013 %Sevilla, Spain and 
%Instituto Carlos I de Fisica Teorica y Computacional,
%Universidad de Granada, E-18071 Granada, Spain.}

%\author[2] {J. Segovia}
%\affiliation[2]{Departamento de Sistemas Fisicos, Quimicos y Naturales, Universidad Pablo de Olavide, E-41013 Sevilla, Spain. }

\begin{abstract}
The LHCb collaboration amplitude analysis of the decays $B^+ \to D^+ D^- K^+$, $B^+ \to D^- D_s^+ \pi^+$, and $B^0 \to \bar{D}^0 D_s^+ \pi^-$ suggested the existence of two new resonant $J^P=0^+$ states with minimum quark content $\bar{u} \bar{d} s c$ and $\bar{s} \bar{u} d c$/$\bar{s} \bar{d} u c$, 
named $T_{cs0}$ and $T_{c\bar{s}0}$, 
respectively. 
In this work, we used the diffusion Monte Carlo (DMC) method to compute the masses of those tetraquarks within the framework of the constituent quark model.  We describe the systems as  compact structures,  not as diquark/antidiquark or meson/meson arrangements. 
In this context, compact means that we used directly the eigenvectors of the spin, color and flavor operators without any splitting and/or (re)combination of smaller units. 
This allows us to deduce the existence of two distinct bound configurations for each composition,  corresponding to ground and excited flavor states.  In both cases,  the excited flavor states are the ones with masses comparable to the ones  observed by LHCb.  In addition,  our analysis allows us to assign unambiguously an isospin value of $I$=1 to the experimentally obtained $T_{cs0}$ tetraquark. \end{abstract}


\end{frontmatter}

%\tableofcontents



\section{Introduction}
\label{introduction}
In 1964, Murray Gell-Mann~\cite{GellMann:1964nj} and George Zweig~\cite{Zweig:1964CERN} independently introduced a highly successful scheme for classifying hadrons based on their valence quark and antiquark content. This framework distinguishes hadrons as either mesons, comprising quark-antiquark pairs, or baryons, formed from three quarks, organized into multiplets according to flavor symmetries.  However, over the past twenty years numerous hadrons have been discovered that are neither mesons not baryons. Predominantly located in the heavy quark sector, these particles are collectively referred to as \emph{XYZ} states. A significant theoretical effort has been undertaken to understand the nature of these exotic hadrons, employing a broad range of approaches. As a result, the literature now contains several comprehensive reviews on the topic~\cite{Dong:2020hxe, Chen:2016spr, Guo:2017jvc, Liu:2019zoy, Yang:2020atz, Dong:2021bvy, Chen:2021erj, Mai:2022eur, Chen:2022asf, Ortega:2020tng}.

In late 2020, the LHCb collaboration performed an amplitude analysis of the decay process $B^+ \to D^+ D^- K^+$, using proton-proton collision data collected at center-of-mass energies of $\sqrt{s} = 7$, $8$, and $13$ TeV, corresponding to a total integrated luminosity of $9\,\text{fb}^{-1}$~\cite{LHCb:2020pxc, LHCb:2020bls}.  That analysis suggested the existence of a new 
tetraquark,  $T_{c s0}(2870)^0$, with 
minimum quark content $\bar{u} \bar{d} sc$,  $J^P=0^+$,  and no assigned value for its isospin.  Its average experimental mass is 
2872 $\pm$ 16 MeV \cite{pdg2024}.

On the other hand,  in December 2022, using again proton-proton collision data with the same parameters, the LHCb collaboration presented a joint amplitude analysis of the decay modes 
$B^0 \to \bar{D}^0 D_s^+ \pi^-$ 
and $B^+ \to D^- D_s^+ \pi^+$
~\cite{LHCb:2022sfr, LHCb:2022lzp}. 
An enhancement observed in the $D_s^+ \pi^+$ invariant mass spectrum suggests the existence of a couple of different open-charm tetraquarks, 
$T_{c\bar{s}0} (2900)$, with minimum quark content $\bar{s}\bar{d} u c$/$\bar{s}\bar{u}d c$.  Both $T_{c\bar{s}0}(2900)$ candidates were assigned isospin $I$=1 and spin-parity quantum numbers $J^P = 0^+$.  The Breit-Wigner parameters of these new resonant states are:
\begin{align}
T_{c\bar s0}^a(2900)^0: \hspace*{0.2cm} M &= (2892 \pm 14 \pm 15) \,\text{MeV} \,, \nonumber \\ \Gamma &= (119 \pm 26 \pm 13) \,\text{MeV} \,, \label{eq:Tcbars0} \\[1ex] \nonumber
T_{c\bar s0}^a(2900)^{++}: \hspace*{0.2cm} M &= (2921 \pm 17 \pm 20) \,\text{MeV} \,, \nonumber \\ \Gamma &= (137 \pm 32 \pm 17) \,\text{MeV} \label{eq:Tcbars++} \,.
\end{align}
with superscripts corresponding to their charges. 
These LHCb collaboration's announcements sparked significant theoretical interest, leading to a wide array of studies on the $T_{cs0}$ candidate employing diverse methodologies: application of QCD sum rules~\cite{Chen:2020aos, Agaev:2020nrc, Zhang:2020oze, Albuquerque:2020ugi, Wang:2020xyc}, non-relativistic and (extended) relativized quark models incorporating various quark-(anti)quark interactions~\cite{Cheng:2020nho, Liu:2020nil, He:2020btl, Xue:2020vtq, He:2020jna, Wang:2020prk, Garcilazo:2020bgc, Karliner:2020vsi, Yang:2021izl, Tan:2020cpu, Hu:2020mxp, Ortega:2023azl, chen2},  effective field theories~\cite{Molina:2010tx, Molina:2020hde, Dong:2020rgs}, and alternative explanations based on triangle singularities~\cite{Liu:2020orv, Burns:2020epm}. Furthermore, various studies have addressed their production and decay properties~\cite{Xiao:2020ltm, Chen:2020eyu, Burns:2020xne,chen3}. In contrast, theoretical work on the $T_{c\bar{s}0}$ signals is currently more limited; for some representative examples, see Refs.~\cite{Wei:2022wtr, Yue:2022mnf, Molina:2022jcd, Yang:2023evp, Ortega:2023azl}.

In this manuscript we analyze the nature of the $T_{c\bar s0}$ and $T_{cs0}$ within the framework of a non-relativistic quark model.  %Even tough we will show some structural properties, we will be 
%focussed mainly in the deduction of the masses of the two tetraquarks considered.  
To solve the corresponding Schrodinger equations, we used the diffusion Monte Carlo (DMC) method,  describing the tetraquarks as compact structures.   
%This approach allows us 
%to 
%account for 
%reduce the uncertainty of the numerical calculation and accounts for 
%the multi-particle correlations in the physical observables.
%The potential model used~\cite{Semay:1994ht, Silvestre-Brac:1996myf} includes pairwise interactions of the most general and accepted type: Coulomb$\,+\,$linear-confining$\,+\,$hyperfine spin-spin.  
%In this work, we describe the tetraquark as a compact structure, what
This implies that we used the full spin-flavor-color function, without resorting to any diquark-antidiquark, antidiquark + light quark+ heavy quark,  or meson/meson picture as in all previous literature.  This allows us to consider both the ground and excited flavor functions for the entire 
system,  with the correct values for the isospin of the full tetraquark. This also implies that our results are not directly comparable to the previous theoretical ones, since the wavefunctions used are not products of those of smaller units. 

The manuscript is organized as follows. After this introduction, the theoretical framework is briefly presented in Sec~\ref{sec:theory}. Section~\ref{sec:results} is mainly devoted to the analysis and discussion of our theoretical results. Finally, we summarize and draw some conclusions in Sec.~\ref{sec:summary}.

\section{Formalism}
\label{sec:theory} 

%\subsection{Subsection title}
Within the framework of the quark model,  a system containing $N$ quarks (in our case,  $N$=4) can be described by a % non-relativistic  
Hamiltonian of the form:
\begin{equation}
H = \sum_{i=1}^N\left(m_i+\frac{p^{\,2}_i}{2m_i}\right) - T_{\text{CM}} + \sum_{i<j}^N V(r_{ij}) \,,
\label{eq:Hamiltonian}
\end{equation}
where $m_{i}$ is the mass of quark $i$,  $\vec{p}_i$ its momentum, and $T_{\text{CM}}$ corresponds to the kinetic energy of the center of mass of the tetraquark.  $V(r_{ij})$ is a phenomenological two-body potential whose
defining parameters are obtained by a fit to experimental properties of mesons and baryons.   In this work, we used the so-called AL1 potential as defined in Refs.  \cite{Semay:1994ht, Silvestre-Brac:1996myf}:
\begin{equation}
V(\vec{r}_{ij}) = V_{\text{OGE}}(r_{ij}) + V_{\text{CON}}(r_{ij})\,.
\end{equation}
The one-gluon exchange (OGE) potential is given by:
\begin{align}
V_{\text{OGE}}(r_{ij}) &= \frac{1}{4} \alpha_{s} \vec{\lambda_{i}}\cdot
\vec{\lambda_{j}}) \Bigg[\frac{1}{r_{ij}} \nonumber \\
&
- \frac{2\pi}{3m_{i}m_{j}} \delta^{(3)}(r_{ij}) \vec{\sigma_{i}}\cdot \vec{\sigma_{j}}) \Bigg] \,,
\label{eq:OGE}
\end{align}
where $\alpha_s$ is the strong coupling constant, $\vec{\lambda_i}$ are the set of SU(3)-{\em color} Gell-Mann matrices, $\vec{\sigma_i}$ denote the Pauli spin matrices,  and the $\delta^{(3)}(r_{ij})$, as customary, is replaced by:
\begin{equation}
\delta^{(3)}(r_{ij}) \to \kappa \, \frac{e^{-r_{ij}^2/r_0^2}}{\pi^{3/2}r_{0}^3} \,,
\end{equation}
with $r_0 = A \left(\frac{2m_im_j}{m_i+m_j}\right)^{-B}$. On the other hand,  the expression for the confining part of the potential is: 
\begin{equation}
V_{\text{CON}}(r_{ij}) = (b\, r_{ij} + \Delta) (\vec{\lambda_{i}}\cdot
\vec{\lambda_{j}}) \,,
\label{eq:CON}
\end{equation}
where $b$ is the confinement strength and $\Delta$ is a global constant fixing the origin of energies. All the parameters in the above expressions for the AL1 potential are given in Table~\ref{tab:parameters}. 

This non-relativistic model has been extensively used in the literature, and its accuracy can be checked in the  benchmarks done in Refs. 
\cite{Silvestre-Brac:1996myf,vijande,baryons} for baryons,  and \cite{mesons} for mesons,  against the experimental masses for systems with light quarks. In this last work there is also a comparison between the masses of mesons obtained with the AL1 potential and its SLM counterpart \cite{jorgeSLM}, that includes terms in the Hamiltonian 
depending on the SU(3)-{\em flavor} Gell-Mann matrices. Since the  the predictions of the SLM potential are not better that those of other non-chiral models, we introduced the flavor 
degree of freedom  exclusively via the symmetry of the total wave function. 

\begin{table}
\caption{\label{tab:parameters} Quark model parameters used in this work and taken from Ref.~\cite{Silvestre-Brac:1996myf}.}
%\begin{ruledtabular}
\begin{center}

\begin{tabular}{llc}
\hline
Quark masses & $m_u=m_d$ (GeV) & 0.315 \\
             & $m_s$ (GeV) & 0.577 \\
             & $m_c$ (GeV) & 1.836 \\
\hline
OGE          & $\alpha_s$        & 0.3802 \\
             & $\kappa$          & 3.6711 \\
             & $A$ (GeV)$^{B-1}$ & 1.6553 \\
             & $B$               & 0.2204 \\
\hline
CON          & $b$ (GeV$^2$)  &  0.1653 \\
             & $\Delta$ (GeV) & -0.8321 \\
\hline
\end{tabular}
\end{center}
%\end{ruledtabular}
\end{table}

To solve the Schrodinger equation derived from the Hamiltonian in Eq. \ref{eq:Hamiltonian}, we used the 
diffusion Monte Carlo (DMC) method \cite{boro94} (in fact, the fixed-node DMC method).  To do so, we write that equation as a standard diffusion equation in imaginary time, $t$
($\hbar=c=1$):
\begin{equation}
-\frac{\partial \Psi_{\alpha'}(\bf{R},t)}{\partial t} = (H_{\alpha'\alpha}-E_s) \Psi_{\alpha}(\bf{R},t) \,,
\label{eq:Sch1}
\end{equation}
where $E_s$ is the usual energy shift used in DMC methods, $\bf{R}\equiv(\vec{r}_1,\ldots,\vec{r}_N)$ stands for the position of the $N$ particles and $\alpha$ denotes each possible spin-flavor-color combination with {\em defined quantum numbers}. The function $\Psi_{\alpha}(\bf{R},t)$ can be expanded in terms of a complete set of the Hamiltonian's eigenfunctions as
\begin{equation}
\Psi_{\alpha}(\bf{R},t) = \sum_i c_{i,\alpha} \, e^{-(E_i-E_s)t} \, \Phi_{i,\alpha}(\bf{R}) \,,
\end{equation}
where the $E_i$ are the eigenvalues of the system's Hamiltonian operator. The ground state wave function, $\phi_{0,\alpha}(\bf{R})$, is obtained as the asymptotic solution of Eq.~\eqref{eq:Sch1} when $t\to \infty$. 
%, as long as there is overlap between $\Psi_{\alpha}(\bf{R},t=0)$ and $\phi_{0,\alpha}(\bf{R})$, 
for any $\alpha$-combination.  This technique has been successfully used in several multiquarks sets 
 \cite{Bai:2016int, Gordillo:2020sgc,Gordillo:2021bra, Ma:2022vqf, Gordillo:2022nnj, Ma:2023int, Mutuk:2023oyz, Gordillo:2023tnz, Alcaraz-Pelegrina:2022fsi, Gordillo:2024sem, Gordillo:2024blx}.
 
The DMC method needs an initial approximation to the many-body wave function of the multiquark, the so-called trial function. That function should include all the information known \emph{a priori} about the system.  In accordance with the previous literature,  we used the expression:
\begin{align}
\Phi_{i,\alpha}(\bf{R}) &\equiv \Phi_i(\vec{r}_1,\ldots,\vec{r}_N;s_1,\ldots,s_N;f_1,\ldots,f_{N_s};c_1,\ldots,c_N) \nonumber \\
&
= \phi_i (\vec{r}_1,\ldots,\vec{r}_N) \times \big[ \chi_s (s_1,\ldots,s_N)  \nonumber \\ 
&
\otimes \chi_f(f_1,\ldots,f_{N_s}) \otimes \chi_c (c_1,\ldots,c_N) \big] \,,
\label{trial}
\end{align}
where, explicitly, $\vec{r}_j$, $s_j$, $f_j$ and $c_j$ stand for the position, spin, flavor and color of the $j$-quark which is inside the $N$-quark cluster.  For this trial function to be a valid, it has to be built as a product of eigenfunctions of the $L^2$ (angular momentum), $C^2$ (color),  $S^2$ (spin) and $F^2$ (flavor) operators for the particular set of eigenvalues considered.   Here,  $N_s$ is the number of light quarks, i.e., those with a flavor degree of freedom.  In the case of the tetraquarks considered here, $N_s$=3,  excluding the $c$ quark. 

In this work, we deal only with hadron states with $L^2$=0,  corresponding to the experimental data, 0$^+$.  This implies that $\phi$ should depend only on the distances between pairs of quarks, being in this way symmetric with respect to the exchange of those positions.  We used the form \cite{Gordillo:2020sgc}:
\begin{equation}
\phi (\vec{r}_1,\ldots,\vec{r}_N) = \prod_{j>i=1}^{N} \exp(-a_{ij} r_{ij}) \,.
\label{eq:radialwf}
\end{equation} 
The constants $a_{ij}$ are chosen to fulfill the cusp conditions derived from the Coulomb-like part of the potential for any pair of quarks.  %In principle, one could consider any other function to describe the position-dependent part of the trial function. In particular, explicit correlations between three and four-body terms could be thought to improve the quality of the approximation.  
%That is what one would expect in a variational calculation. However,  
%Since DMC is not a variational method,  
The DMC algorithm is able to correct the possible shortcomings of Eq. \ref{eq:radialwf} to produce the real position-dependent many-body function of the multiquark, since it is not a variational method \cite{Hammond:1994bk}. 
This means that any initial guess with a {\em non-negligible overlap} with the real ground state will produce the exact solution for $t\to \infty$.  In particular, it has been proved
\cite{boro94}, that  the introduction of three-body correlations is crucial to improve the results of variational calculations, but it is irrelevant (within statistical errors) when we use DMC.   Conversely, if we use a trial function that we know a priori to be completely orthogonal to the ground state, the DMC technique converges to the corresponding  excited state \cite{Hammond:1994bk}. This is routinely done in the field of cold gases, in which trial functions corresponding to scattering states are used. 

The trial funcion contains also contributions from the spin, flavor and color degrees of freedom  ($\chi_s$, $\chi_f$ and $\chi_c$).  In the same way that $\phi$ is an eigenvalue of 
$L^2$,  all the $\chi$'s  have to be eigenvectors of the spin, flavor and color operators, respectively, defined by:
\begin{eqnarray}
S^2 = \left(\sum_{i=1}^{N} \frac{\vec{\sigma_i}}{2} \right)^2  \nonumber
F^2 = \left(\sum_{i=1}^{N_s} \frac{\vec{\lambda_i^f}}{2} \right)^2  \nonumber
C^2 = \left(\sum_{i=1}^{N} \frac{\vec{\lambda_i}}{2} \right)^2,
\label{eq:operators}
\end{eqnarray}
with fixed eigenvalues.  $\vec{\lambda^f}$ is the corresponding set of Gell-Mann matrices  representing the {\em  flavor} degrees of freedom $u$, $d$ and $s$ instead of the color values $r$, $g$ and $b$
~\cite{libro2018}.  Since we aim to compare our results with the experimental ones,  we fixed $S^2$=0 (corresponding to a 0$^+$ state) and,  obviously,  $C^2$=0 (colorless wavefunction).

On the other hand, 
%the flavor operator, $F^2$ has other constraints.  In principle, 
we can choose any flavor eigenvector to describe the system, not necessarily the one corresponding to the lowest eigenvalue.  The only limitation is that those functions have to be {\em simultaneously} eigenvectors of the isospin operator, defined by:
\begin{equation}
I^2 = \left(\sum_{i=1}^{N_s} \frac{\vec{\lambda_i^f}}{2} \right)^2, 
\end{equation}
%isomorphic with the spin operator when we have only $u$ and $d$ quarks, i.e., we consider only the 
that includes only the $\lambda^{f,1}_i, \lambda^{f,2}_i$ and $\lambda^{f,3}_i$ components of the $\vec{\lambda^f}_i$ set, and it is isomorphic with the spin operator when we have only $u$ and $d$ quarks. This means that we have to obtain first the eigenvectors of $I^2$, and then solve for the eigenfunctions of $F^2$ in that constrained space.  

We are considering only tetraquarks
with composition $\bar{u} \bar{d} s c$, with electric charge equal to 0,  and $\bar{s} \bar{u} d c$ (identical,  in our formalism, to $\bar{s} \bar{d} u c$), with electric charge equal to +2 (0 for $\bar{s} \bar{d} u c$), in order to compare our results with experiments.  Those values for the total charges rule out the compositions $\bar{s} \bar{n} n c$ and $\bar{n} \bar{n} s c$ and with $n=u,d$. 
%This means our results are not directly comparable to those of Refs. \cite{chen2,Ortega:2023azl}. 
In the first case,  corresponding to the $T_{c\bar{s}0}(2900)$ system,  the only flavor eigenfunctions of the $I^2$ operator (without the $c$ quark) are:
\begin{equation}
|\bar s\bar u d \rangle \quad\text{and}\quad |\bar u \bar s d \rangle \,,
\end{equation}
with eigenvalue $I$=1, in accordance with the experimental data.  However,  they are {\em not} eigenvectors of the {\em total} flavor operator, $F^2$, and cannot be used as such in the calculations.   The proper eigenfunctions of the complete flavor operator are:
\begin{align}
|\chi_{f1}\rangle &= \frac{1}{\sqrt{2}} \left(|\bar s\bar u d\rangle - |\bar u \bar s d \rangle\right) \,, \\
|\chi_{f2}\rangle &= \frac{1}{\sqrt{2}} \left(|\bar s\bar u d\rangle + |\bar u \bar s d \rangle\right) \,,
\end{align}
with eigenvalues 10/3 and 16/3, respectively.  Only those functions can be used in Eq. \ref{trial} to build the total wavefunction describing the tetraquark. Repeating the procedure for  $\bar{u} \bar{d} s c$
($T_{cs0} (2870)^0$ tetraquark),  we have two eigenvectors for the $I^2$ operator:
\begin{align}
|\chi_{f1}^\prime\rangle &= \frac{1}{\sqrt{2}} \left(|\bar u\bar d s\rangle - |\bar d \bar u s \rangle\right) \,, \\
|\chi_{f2}^\prime\rangle &= \frac{1}{\sqrt{2}} \left(|\bar u\bar d s\rangle + |\bar d \bar u s \rangle\right) \,.
\end{align}
the first one with eigenvalue $I$=0, and the second one with value $I$=1.  Each function is, at the same time,  an eigenvector of the flavor operator $F^2$, with eigenvalues 10/3 and 16/3, respectively.  
%We have to introduce each of those functions in Eq. \ref{trial} to describe the different tetraquarks considered.  This set of functions correspond to the $T_{cs0} (2870)^0$ tetraquark.  
$\chi_{f1}$ and $\chi_{f1}^\prime$ are antisymmetric with respect to the exchange of two antiquarks ($u$, $d$ and $s$ being undistinguishable under the flavor operator), while $\chi_{f2}$ and $\chi_{f2}^\prime$ are symmetrical with respect to the same 
exchange. 

However, the trial function 
%the description of the arrangements is not complete, since we have to consider the fermionic nature of the quarks.  
%This implies that Eq. \eqref{trial} 
has to be antisymmetric with respect to the exchange of any two 
identical quarks, since quarks are fermions that have to comply with Pauli statistics.  The most general way to produce 
adequate 
%spin-flavor-color 
combinations is to 
%Since Eq.~\eqref{eq:radialwf} is symmetric with respect to those changes,  we have to produce flavor-color-spin combinations which are antisymmetric to comply with Pauli statistics. To do so, we 
apply the antisymmetry operator,
\begin{equation}
\mathcal{A} = \frac{1}{N_p} \sum_{{\alpha}=1}^N (-1)^P \mathcal{P_{\alpha}} \,,
\label{eq:antisymope}
\end{equation}
to the complete set of spin-flavor-color functions.  In Eq.~\eqref{eq:antisymope}, $N_p$ is the number of possible permutations of the set of quark indexes, $P$ is the order of the permutation, and $\mathcal{P_{\alpha}}$ represents the matrices that define those permutations. 
Once constructed the matrix derived from the operator in Eq.~\eqref{eq:antisymope}, we have to check if we can find any eigenvector with eigenvalue equal to one.  Those combinations are the input of the DMC calculation. 

That is the most general approach and can be applied to large multiquarks \cite{Gordillo:2023tnz}.  However, for small systems as the ones considered in this work,  we could naively  obtain the eigenvectors of the color and spin operators and combine them with the flavor functions 
%to produce wavefunctions of the proper symmetry 
by hand.  The colorless functions are \cite{Gordillo:2020sgc}:
\begin{eqnarray}
|\bar{3}_{12} 3_{34}\rangle = \frac{1}{\sqrt{12}} \Big( 
%&
+|\bar{r}\bar{g}rg\rangle
+|\bar{g}\bar{r}gr\rangle
-|\bar{g}\bar{r}rg\rangle \nonumber \\
%&
-|\bar{r}\bar{g}gr\rangle
+|\bar{r}\bar{b}rb\rangle   
+|\bar{b}\bar{r}br\rangle %\nonumber \\
%&
-|\bar{b}\bar{r}rb\rangle
-|\bar{r}\bar{b}br\rangle
+|\bar{g}\bar{b}gb\rangle \nonumber \\
%&
+|\bar{b}\bar{g}bg\rangle
-|\bar{b}\bar{g}gb\rangle
-|\bar{g}\bar{b}bg\rangle \Big) \,,
\end{eqnarray}
which is antisymmetric under the exchange of either both quarks or both antiquarks, and
\begin{eqnarray}
|\bar{6}_{12} 6_{34}\rangle = \frac{1}{\sqrt{6}} \Big[
%&
+|\bar{r}\bar{r}rr\rangle 
+|\bar{g}\bar{g}gg\rangle
+|\bar{b}\bar{b}bb\rangle  \nonumber \\
%&
+\frac{1}{2} \Big( 
+|\bar{r}\bar{g}rg\rangle 
+|\bar{g}\bar{r}gr\rangle 
+|\bar{g}\bar{r}rg\rangle %\nonumber \\
%& 
%\hspace*{0.90cm} 
+|\bar{r}\bar{g}gr\rangle
+|\bar{r}\bar{b}rb\rangle
+|\bar{b}\bar{r}br\rangle \nonumber \\
%&
%\hspace*{0.90cm} 
+|\bar{b}\bar{r}rb\rangle
+|\bar{r}\bar{b}br\rangle
+|\bar{g}\bar{b}gb\rangle%\nonumber \\
%&
%\hspace*{0.90cm} 
+|\bar{b}\bar{g}bg\rangle
+|\bar{b}\bar{g}gb\rangle
+|\bar{g}\bar{b}bg\rangle \Big) \Big] \,,
\end{eqnarray}
which is symmetric under the exchange of either both quarks or both antiquarks.   On the other hand,   the two linearly independent wavefunctions with spin 0 are:
%\begin{subequations}
\begin{align}
|\chi_{S=0,S_z=0}\rangle_{\text{SS}} &= \frac{1}{\sqrt{12}} \Big( +2\,|\downarrow\downarrow\uparrow\uparrow\rangle +2\,|\uparrow\uparrow\downarrow\downarrow\rangle \nonumber \\
& 
\hspace*{1.45cm} 
-|\downarrow\uparrow\uparrow\downarrow\rangle -|\uparrow\downarrow\downarrow\uparrow\rangle %\nonumber \\
%&
%\hspace*{1.45cm} 
+|\downarrow\uparrow\downarrow\uparrow\rangle -|\uparrow\downarrow\uparrow\downarrow\rangle \Big) \,, \\
|\chi_{S=0,S_z=0}\rangle_{\text{AA}} &= \frac{1}{2} \Big( -|\downarrow\uparrow\downarrow\uparrow \rangle - |\downarrow\uparrow\uparrow\downarrow \rangle %\nonumber \\
%&
%\hspace*{1.00cm} 
- |\uparrow\downarrow\downarrow\uparrow \rangle + |\uparrow\downarrow\uparrow\downarrow\rangle \Big) \,.
\end{align}
%\end{subequations}
They are, respectively, symmetric ($S$) and antisymmetric ($A$) under the exchange of both quarks and both antiquarks.  This means that for the $T_{c\bar{s}0}(2900)$ system we have two possible flavor-color-spin combinations  for each value of $F^2$: 
 \begin{eqnarray}
\Phi_{1a} = |\chi_{f1} \rangle \otimes  |\bar{6}_{12} 6_{34}\rangle \otimes  |\chi_{S=0,S_z=0}\rangle_{\text{AA}} \nonumber \\
\Phi_{1b} = |\chi_{f1}\rangle  \otimes |\bar{3}_{12} 3_{34}\rangle \otimes |\chi_{S=0,S_z=0}\rangle_{\text{SS}}
\end{eqnarray}
for $F^2$ = 10/3, and  
\begin{eqnarray}
\Phi_{2a} = |\chi_{f2}\rangle  \otimes  |\bar{3}_{12} 3_{34}\rangle \otimes  |\chi_{S=0,S_z=0}\rangle_{\text{SS}} \nonumber \\
\Phi_{2b} = |\chi_{f2}\rangle \otimes  |\bar{6}_{12} 6_{34}\rangle \otimes |\chi_{S=0,S_z=0}\rangle_{\text{AA}} 
\end{eqnarray} 
for $F^2$ = 16/3.  Those sets of functions are antisymmetric with respect to the exchange of two antiquarks and have to be multiplied by Eq. \ref{eq:radialwf}.  A similar set of functions (with primed flavor  parts) describe the $T_{cs0} (2870)^0$ tetraquark.  Each pair can be introduced directly in the DMC algorithm, since they have a defined set of quantum numbers. However, those are not the only possibilities. 
For instance,  instead of $\Phi_{1a}$ and $\Phi_{1b}$ we can use $1/\sqrt(2)$ ($\Phi_{1a}$ + $\Phi_{1b}$) and $1/\sqrt(2)$ ($\Phi_{1a}$ - $\Phi_{1b}$) and obtain identical results, since those combinations are also eigenvalues of the antisymmetry operator and have the same eigenvalues of the flavor, spin and color operators.

\section{Results}
\label{sec:results}

The DMC is a Monte Carlo technique and, as such,  can be affected by several sources of statistical and systematic errors. 
First,  we have an imaginary time-step error derived from application of the short-time approximation \cite{Hammond:1994bk} to the 
real Green function of the system.  This time step should be small enough to make that approximation reliable and large enough to 
ensure that the phase space of the system under consideration is properly sampled.  We checked that 10$^{-6}$ MeV$^{-1}$ (of the same order of magnitude but smaller that the used in Ref.  \cite{baryons}) was enough for both purposes.  With that step, we produced simulation runs 6 $\times$ 10$^5$ steps long,  discarding the first 10$^5$ to ensure equilibration.  The values of the observables for the remaining  5 $\times$ 10$^5$ steps were plotted against the simulation time to ensure that there were no drifts that could produce any systematic deviations.  None were observed.  As an additional test,  each simulation was repeated a minimum of 3 times starting from different configurations to check if the results were comparable. To avoid spurious correlations within each simulation run,  we considered for averaging only configurations separated by 500 steps., i.e.,  we have a sample of 1000 points.  The statistical uncertainties of those measurements were in the range 2-3 MeV.  The same interval contained the averages of the three independent simulation runs.  Those uncertainties were of the same order of magnitude of those of Ref.  \cite{baryons}.  Finally, the number of walkers used for any simulations were 10$^3$.  To increase that number did not change the quality of the simulation results.  

The fermionic nature of the quarks implies that we have nodes in the total wavefunction, nodes circumscribed to the spin-flavor-color part.  However, since we used the exact eigenvectors of those operators, we know a priori the position of those nodes. In that case, the DMC method is also able to converge to the exact state of minimum energy compatible with the symmetry (and hence set of nodes) of the (compact) trial function considered by using a fixed-node DMC algorithm \cite{Hammond:1994bk}.   On the other hand, if the nodes of the trial wavefunction are not the exact ones,  the fixed-node technique is not able to correct their position,  rendering an upper bound of the energy. 

The masses of the $T_{c s0}(2870)$ and $T_{c \bar{s}0}(2900)$ tetraquarks computed in this work are given in Table~\ref{tab:masses}, together with the experimental values.  We also include the masses of all the mesons that can be combined to form the corresponding tetraquarks,  both calculated with the AL1 potential and obtained in experiments \cite{pdg2024}.   The comparison between both series of data serves to attest again the reliability of the AL1 potential in a non-relativistic framework.   In the  $T_{c \bar{s}0}(2900)$ case, we used the experimental combined mass of the tetraquarks since our description is not able to distinguish between $\bar{s} \bar{u} d c$ and $\bar{s} \bar{d} u c$.  What we observe is that our approximation produces masses comparable to those in the experiments only if we consider the tetraquarks in their flavor excited states ($F^2$ = 16/3).   This implies that we should have another set of two tetraquarks in their respective flavor ground states with masses $\sim$400 MeV lower.  This is perfectly possible since,  for example,  we can find the $uds$ baryon in two isospin states, $I$=0 and $I$=1 \cite{libro2018}.  The comparison between our masses and the experimental ones allows us also to assign the value $I$=1 to the isospin of the  $T_{c s0}(2870)$ tetraquark,  something that has not been experimentally done.  It also explains the value $I$=1 for the  $T_{c \bar{s}0}(2900)$ system.  
We can see also that both tetraquarks have masses that are larger or, at most, comparable to those of the proper combination of mesons.  The reason why the DMC algorithm does not converge to those smaller masses is that we have {\em chosen} to  model the system as compact,  not as a set of meson $\times$ meson spin-flavor-color configurations.  Thus, the technique converges to the minimum energy value compatible with that compact function, that has nodes in different positions
than the meson $\times$ meson one.  

\begin{table}[!t]
\caption{\label{tab:masses} Masses of the compact tetraquarks calculated in this work,  together with their closest experimental data.  Also included are the masses of the relevant $S=0$ and $S=1$ mesons that can be coupled to produce $0^+$ tetraquarks and the corresponding direct sums. }
\begin{tabular}{llccc}
%$T_{cs0} (2870)^0$ \\
%\hline
Quark content & $I$ & $F^2$ & Mass (MeV) & Exp.  (MeV)\\
\hline
  $\bar{u} \bar{d} s c$  & 0  & 10/3 & 2594 $\pm$ 2 & \\
                                       & 1  & 16/3 &  2912 $\pm$ 2  & 2872 \\
\hline
  $\bar{s} \bar{u} d c$ / $\bar{s} \bar{d} u c$ & 1  & 10/3 & 2526 $\pm$ 3 & \\
                                       & 1  & 16/3 &  2920 $\pm$ 3 & 2908   \\
\hline
$S=0$  mesons & & & \\
\hline
$\bar{u} d$ / $\bar{d} u$  & & & 142 $\pm$ 2 & 140 \\
$\bar{u} s$ / $\bar{d}s$ &     &    &  492 $\pm$ 2 & 494 / 498 \\                                     
$\bar{u} c$ / $\bar{d} c$ & & & 1861 $\pm$ 2 & 1865 / 1870 \\
$\bar{s} c$ & & & 1961 $\pm$ 2 & 1968 \\
\hline
$\bar{u} s$ + $\bar{d} c$ / $\bar{d} s$ + $\bar{u} c$ & & & 2359 $\pm$ 3 & 2364 / 2363  \\
$\bar{u} d$ + $\bar{s} c$ / $\bar{d} u$ + $\bar{s} c$ & & & 2103 $\pm$ 3 & 2108 \\
\hline
$S=1$  mesons & & & \\
\hline
$\bar{u} d$ / $\bar{d} u$  & & & 771 $\pm$ 2 & 775 \\
$\bar{u} s$ / $\bar{d}s$ &     &    &  904 $\pm$ 2 & 892/ 896  \\                                     
$\bar{u} c$ / $\bar{d} c$ & & & 2017 $\pm$ 2 & 2007 / 2010 \\
$\bar{s} c$ & & & 2102  $\pm$ 2 & 2112 \\
\hline
$\bar{u} s$ + $\bar{d} c$ / $\bar{d} s$ + $\bar{u} c$ & & & 2921 $\pm$ 3 & 2902 / 2903  \\
$\bar{u} d$ + $\bar{s} c$ / $\bar{d} u$ + $\bar{s} c$ & & & 2873 $\pm$ 3 & 2887 \\
\end{tabular}
\end{table}

\begin{figure*}
\includegraphics[width=0.49\textwidth]{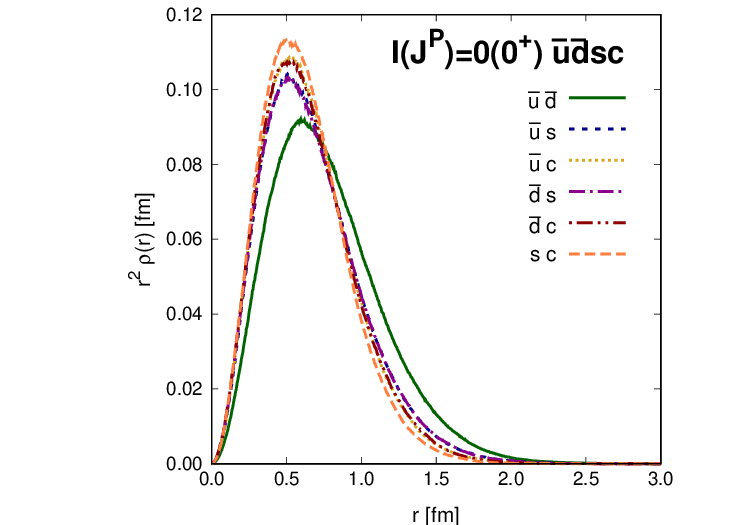}
\includegraphics[width=0.49\textwidth]{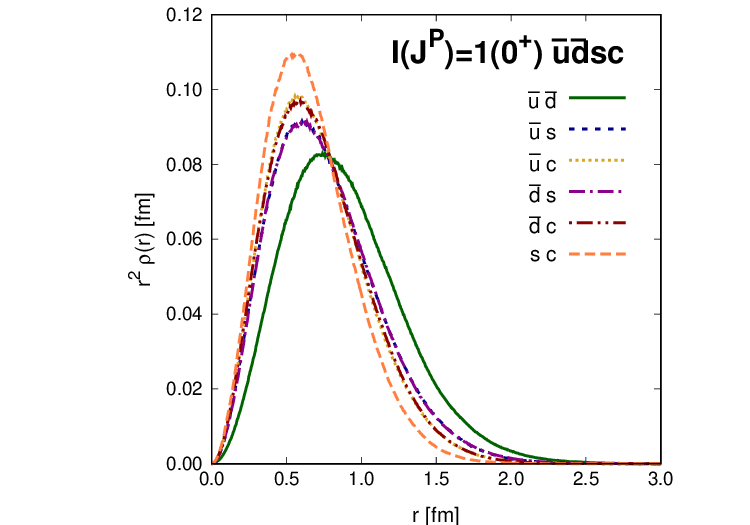} \\ [1ex]
\includegraphics[width=0.49\textwidth]{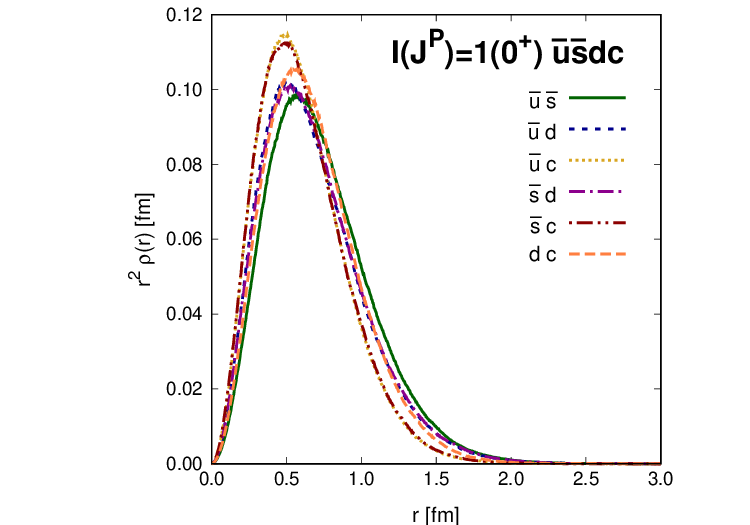}
\includegraphics[width=0.49\textwidth]{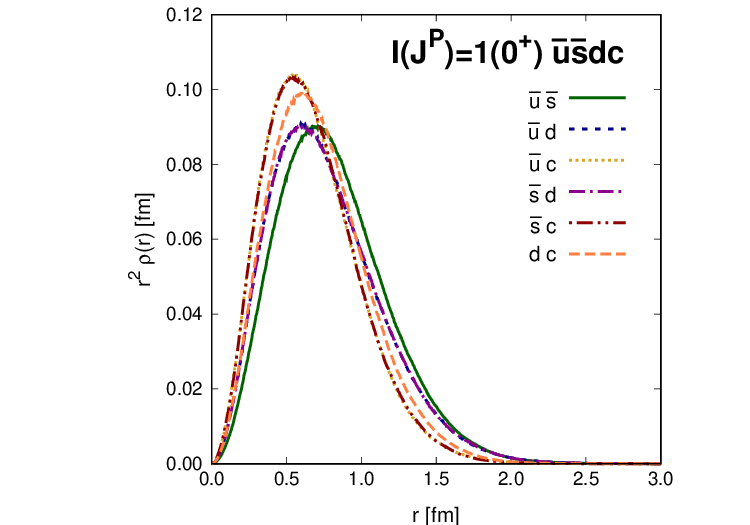}
\vspace{1cm}
\caption{Radial distribution functions of the ground and excited flavor states of the $T_{cs0}$ (top row) and $T_{c\bar s0}$ (bottom row) tetraquarks. These functions represent the probability of finding two different kinds of quarks (antiquarks) at an interquark distance $r$.  Bottom left,  $F^2$ = 10/3; Bottom right, $F^2$ = 16/3. }
\label{fig:Correlations} 
\end{figure*}

We turn now our attention on the structure of the tetraquarks,  using the radial distribution functions.  Those give us the probability of having a quark of a particular type at a given distance from another one located at the origin of coordinates \cite{Gordillo:2020sgc}.  Figure~\ref{fig:Correlations} shows the results for all the quark pairs in all the tetraquarks considered in this work.  One thing is immediately apparent: all the two-body functions are very similar to each other, with peaks in the 0.5-1 fm range.  This implies that the tetraquarks are compact structures with no obvious splitting into smaller structures such as molecules or diquark-antidiquark arrangements.  In those cases, we would expect two radial functions with maxima well separated from each other corresponding to the $\bar{q} q$+$\bar{q'} q'$ mesons (or $qq$+$\bar{q'} \bar{q'}$ counterparts), and a different set of functions for the cross-correlation between quarks in different subunits. This is what can be seen in the molecular pattern found in Ref.  \cite{Gordillo:2021bra} for the $X(3872)$ tetraquark.  No such such three-function splitting is seen in Fig.  \ref{fig:Correlations}.  

%%%%%%%%%%%%%%%%%%%%%%%%%%%%%%%%%%%%%%%%%%%%%%%%%%%%%%%%%%%%%%%%%%%%%%%%%%%%%%%%

\section{CONCLUSIONS}
\label{sec:summary}

We have explored the nature of recently observed $T_{cs0}$ and $T_{c\bar{s}0}$ tetraquark candidates, using the diffusion Monte Carlo method applied to a non-relativistic quark model.  A comparison of our results to the experimental data available indicates that the resonances observed are in both cases excited flavor states with $I$=1.  This implies the existence of their ground state flavor counterparts that could be targeted in future experimental searches.  We have verified also that all systems are compact structures, with no obvious splitting into meson/meson or quark/antidiquark subunits.  
%%\label{}
%\lipsum[4]

\section*{Acknowledgements}
We acknowledge financial support from 
Ministerio Espa\~nol de Ciencia e Innovaci\'on under grant Nos. PID2023-147469NB-C21 and PID2022-140440NB-C22;
Junta de Andaluc\'ia under contract Nos. PAIDI FQM-205 and FQM-370 as well as PCI+D+i under the title: ``Tecnolog\'\i as avanzadas para la exploraci\'on del universo y sus componentes" (Code AST22-0001).
%

%\appendix

%\section{Appendix title 1}

%\section{Appendix title 2}

\bibliographystyle{elsarticle-harv} 
\bibliography{print_DMC-Tcs}

\begin{thebibliography}{68}
\expandafter\ifx\csname natexlab\endcsname\relax\def\natexlab#1{#1}\fi
\providecommand{\url}[1]{\texttt{#1}}
\providecommand{\href}[2]{#2}
\providecommand{\path}[1]{#1}
\providecommand{\DOIprefix}{doi:}
\providecommand{\ArXivprefix}{arXiv:}
\providecommand{\URLprefix}{URL: }
\providecommand{\Pubmedprefix}{pmid:}
\providecommand{\doi}[1]{\href{http://dx.doi.org/#1}{\path{#1}}}
\providecommand{\Pubmed}[1]{\href{pmid:#1}{\path{#1}}}
\providecommand{\bibinfo}[2]{#2}
\ifx\xfnm\relax \def\xfnm[#1]{\unskip,\space#1}\fi
%Type = Article
\bibitem[{Aaij et~al.(2020a)}]{LHCb:2020bls}
\bibinfo{author}{Aaij, R.}, et~al. (\bibinfo{collaboration}{LHCb}),
  \bibinfo{year}{2020}a.
\newblock \bibinfo{title}{{A model-independent study of resonant structure in
  $B^+\to D^+D^-K^+$ decays}}.
\newblock \bibinfo{journal}{Phys. Rev. Lett.} \bibinfo{volume}{125},
  \bibinfo{pages}{242001}.
\newblock \DOIprefix\doi{10.1103/PhysRevLett.125.242001},
  \href{http://arxiv.org/abs/2009.00025}{{\tt arXiv:2009.00025}}.
%Type = Article
\bibitem[{Aaij et~al.(2020b)}]{LHCb:2020pxc}
\bibinfo{author}{Aaij, R.}, et~al. (\bibinfo{collaboration}{LHCb}),
  \bibinfo{year}{2020}b.
\newblock \bibinfo{title}{{Amplitude analysis of the $B^+\to D^+D^-K^+$
  decay}}.
\newblock \bibinfo{journal}{Phys. Rev. D} \bibinfo{volume}{102},
  \bibinfo{pages}{112003}.
\newblock \DOIprefix\doi{10.1103/PhysRevD.102.112003},
  \href{http://arxiv.org/abs/2009.00026}{{\tt arXiv:2009.00026}}.
%Type = Article
\bibitem[{Aaij et~al.(2023a)}]{LHCb:2022lzp}
\bibinfo{author}{Aaij, R.}, et~al. (\bibinfo{collaboration}{LHCb}),
  \bibinfo{year}{2023}a.
\newblock \bibinfo{title}{{Amplitude analysis of
  B0\textrightarrow{}D\textasciimacron{}0Ds+\ensuremath{\pi}- and
  B+\textrightarrow{}D-Ds+\ensuremath{\pi}+ decays}}.
\newblock \bibinfo{journal}{Phys. Rev. D} \bibinfo{volume}{108},
  \bibinfo{pages}{012017}.
\newblock \DOIprefix\doi{10.1103/PhysRevD.108.012017},
  \href{http://arxiv.org/abs/2212.02717}{{\tt arXiv:2212.02717}}.
%Type = Article
\bibitem[{Aaij et~al.(2023b)}]{LHCb:2022sfr}
\bibinfo{author}{Aaij, R.}, et~al. (\bibinfo{collaboration}{LHCb}),
  \bibinfo{year}{2023}b.
\newblock \bibinfo{title}{{First Observation of a Doubly Charged Tetraquark and
  Its Neutral Partner}}.
\newblock \bibinfo{journal}{Phys. Rev. Lett.} \bibinfo{volume}{131},
  \bibinfo{pages}{041902}.
\newblock \DOIprefix\doi{10.1103/PhysRevLett.131.041902},
  \href{http://arxiv.org/abs/2212.02716}{{\tt arXiv:2212.02716}}.
%Type = Article
\bibitem[{Agaev et~al.(2021)Agaev, Azizi and Sundu}]{Agaev:2020nrc}
\bibinfo{author}{Agaev, S.S.}, \bibinfo{author}{Azizi, K.},
  \bibinfo{author}{Sundu, H.}, \bibinfo{year}{2021}.
\newblock \bibinfo{title}{{New scalar resonance X 0(2900) as a molecule: mass
  and width}}.
\newblock \bibinfo{journal}{J. Phys. G} \bibinfo{volume}{48},
  \bibinfo{pages}{085012}.
\newblock \DOIprefix\doi{10.1088/1361-6471/ac0b31},
  \href{http://arxiv.org/abs/2008.13027}{{\tt arXiv:2008.13027}}.
%Type = Article
\bibitem[{Albuquerque et~al.(2021)Albuquerque, Narison, Rabetiarivony and
  Randriamanatrika}]{Albuquerque:2020ugi}
\bibinfo{author}{Albuquerque, R.M.}, \bibinfo{author}{Narison, S.},
  \bibinfo{author}{Rabetiarivony, D.}, \bibinfo{author}{Randriamanatrika, G.},
  \bibinfo{year}{2021}.
\newblock \bibinfo{title}{{$X_{0,1}$(2900) and $(D^-K^+)$ invariant mass from
  QCD Laplace sum rules at NLO}}.
\newblock \bibinfo{journal}{Nucl. Phys. A} \bibinfo{volume}{1007},
  \bibinfo{pages}{122113}.
\newblock \DOIprefix\doi{10.1016/j.nuclphysa.2020.122113},
  \href{http://arxiv.org/abs/2008.13463}{{\tt arXiv:2008.13463}}.
%Type = Article
\bibitem[{Alcaraz-Pelegrina and Gordillo(2022)}]{Alcaraz-Pelegrina:2022fsi}
\bibinfo{author}{Alcaraz-Pelegrina, J.M.}, \bibinfo{author}{Gordillo, M.C.},
  \bibinfo{year}{2022}.
\newblock \bibinfo{title}{{Diffusion Monte~Carlo calculations of fully heavy
  compact hexaquarks}}.
\newblock \bibinfo{journal}{Phys. Rev. D} \bibinfo{volume}{106},
  \bibinfo{pages}{114028}.
\newblock \DOIprefix\doi{10.1103/PhysRevD.106.114028},
  \href{http://arxiv.org/abs/2205.13886}{{\tt arXiv:2205.13886}}.
%Type = Book
\bibitem[{Amsler(2018)}]{libro2018}
\bibinfo{author}{Amsler, C.}, \bibinfo{year}{2018}.
\newblock \bibinfo{title}{The quark structure of hadrons. Lecture Notes in
  Physics 949}.
\newblock \bibinfo{publisher}{Springer}.
%Type = Article
\bibitem[{Bai et~al.(2019)Bai, Lu and Osborne}]{Bai:2016int}
\bibinfo{author}{Bai, Y.}, \bibinfo{author}{Lu, S.}, \bibinfo{author}{Osborne,
  J.}, \bibinfo{year}{2019}.
\newblock \bibinfo{title}{{Beauty-full Tetraquarks}}.
\newblock \bibinfo{journal}{Phys. Lett. B} \bibinfo{volume}{798},
  \bibinfo{pages}{134930}.
\newblock \DOIprefix\doi{10.1016/j.physletb.2019.134930},
  \href{http://arxiv.org/abs/1612.00012}{{\tt arXiv:1612.00012}}.
%Type = Article
\bibitem[{Boronat and Casulleras(1994)}]{boro94}
\bibinfo{author}{Boronat, J.}, \bibinfo{author}{Casulleras, J.},
  \bibinfo{year}{1994}.
\newblock \bibinfo{title}{{Monte Carlo analysis of an interatomic potential for
  He}}.
\newblock \bibinfo{journal}{Phys. Rev. B} \bibinfo{volume}{49},
  \bibinfo{pages}{8920--8930}.
\newblock \URLprefix \url{https://link.aps.org/doi/10.1103/PhysRevB.49.8920},
  \DOIprefix\doi{10.1103/PhysRevB.49.8920}.
%Type = Article
\bibitem[{Burns and Swanson(2021a)}]{Burns:2020xne}
\bibinfo{author}{Burns, T.J.}, \bibinfo{author}{Swanson, E.S.},
  \bibinfo{year}{2021}a.
\newblock \bibinfo{title}{{Discriminating among interpretations for the
  $X(2900)$ states}}.
\newblock \bibinfo{journal}{Phys. Rev. D} \bibinfo{volume}{103},
  \bibinfo{pages}{014004}.
\newblock \DOIprefix\doi{10.1103/PhysRevD.103.014004},
  \href{http://arxiv.org/abs/2009.05352}{{\tt arXiv:2009.05352}}.
%Type = Article
\bibitem[{Burns and Swanson(2021b)}]{Burns:2020epm}
\bibinfo{author}{Burns, T.J.}, \bibinfo{author}{Swanson, E.S.},
  \bibinfo{year}{2021}b.
\newblock \bibinfo{title}{{Kinematical cusp and resonance interpretations of
  the $X(2900)$}}.
\newblock \bibinfo{journal}{Phys. Lett. B} \bibinfo{volume}{813},
  \bibinfo{pages}{136057}.
\newblock \DOIprefix\doi{10.1016/j.physletb.2020.136057},
  \href{http://arxiv.org/abs/2008.12838}{{\tt arXiv:2008.12838}}.
%Type = Article
\bibitem[{Chen(2022)}]{Chen:2021erj}
\bibinfo{author}{Chen, H.X.}, \bibinfo{year}{2022}.
\newblock \bibinfo{title}{{Hadronic molecules in B decays}}.
\newblock \bibinfo{journal}{Phys. Rev. D} \bibinfo{volume}{105},
  \bibinfo{pages}{094003}.
\newblock \DOIprefix\doi{10.1103/PhysRevD.105.094003},
  \href{http://arxiv.org/abs/2103.08586}{{\tt arXiv:2103.08586}}.
%Type = Article
\bibitem[{Chen et~al.(2020)Chen, Chen, Dong and Su}]{Chen:2020aos}
\bibinfo{author}{Chen, H.X.}, \bibinfo{author}{Chen, W.},
  \bibinfo{author}{Dong, R.R.}, \bibinfo{author}{Su, N.}, \bibinfo{year}{2020}.
\newblock \bibinfo{title}{{$X_0$(2900) and $X_1$(2900): Hadronic Molecules or
  Compact Tetraquarks}}.
\newblock \bibinfo{journal}{Chin. Phys. Lett.} \bibinfo{volume}{37},
  \bibinfo{pages}{101201}.
\newblock \DOIprefix\doi{10.1088/0256-307X/37/10/101201},
  \href{http://arxiv.org/abs/2008.07516}{{\tt arXiv:2008.07516}}.
%Type = Article
\bibitem[{Chen et~al.(2017a)Chen, Chen, Liu, Liu and Zhu}]{Chen:2016spr}
\bibinfo{author}{Chen, H.X.}, \bibinfo{author}{Chen, W.}, \bibinfo{author}{Liu,
  X.}, \bibinfo{author}{Liu, Y.R.}, \bibinfo{author}{Zhu, S.L.},
  \bibinfo{year}{2017}a.
\newblock \bibinfo{title}{{A review of the open charm and open bottom
  systems}}.
\newblock \bibinfo{journal}{Rept. Prog. Phys.} \bibinfo{volume}{80},
  \bibinfo{pages}{076201}.
\newblock \DOIprefix\doi{10.1088/1361-6633/aa6420},
  \href{http://arxiv.org/abs/1609.08928}{{\tt arXiv:1609.08928}}.
%Type = Article
\bibitem[{Chen et~al.(2023)Chen, Chen, Liu, Liu and Zhu}]{Chen:2022asf}
\bibinfo{author}{Chen, H.X.}, \bibinfo{author}{Chen, W.}, \bibinfo{author}{Liu,
  X.}, \bibinfo{author}{Liu, Y.R.}, \bibinfo{author}{Zhu, S.L.},
  \bibinfo{year}{2023}.
\newblock \bibinfo{title}{{An updated review of the new hadron states}}.
\newblock \bibinfo{journal}{Rept. Prog. Phys.} \bibinfo{volume}{86},
  \bibinfo{pages}{026201}.
\newblock \DOIprefix\doi{10.1088/1361-6633/aca3b6},
  \href{http://arxiv.org/abs/2204.02649}{{\tt arXiv:2204.02649}}.
%Type = Article
\bibitem[{Chen et~al.(2017b)Chen, Chen, Liu, Steele and Zhu}]{chen2}
\bibinfo{author}{Chen, W.}, \bibinfo{author}{Chen, H.X.}, \bibinfo{author}{Liu,
  X.}, \bibinfo{author}{Steele, T.G.}, \bibinfo{author}{Zhu, S.L.},
  \bibinfo{year}{2017}b.
\newblock \bibinfo{title}{Open-flavor charm and bottom
  $sq\overline{q}\overline{Q}$ and $qq\overline{q}\overline{Q}$ tetraquark
  states}.
\newblock \bibinfo{journal}{Phys. Rev. D} \bibinfo{volume}{95},
  \bibinfo{pages}{114005}.
\newblock \URLprefix \url{https://link.aps.org/doi/10.1103/PhysRevD.95.114005},
  \DOIprefix\doi{10.1103/PhysRevD.95.114005}.
%Type = Article
\bibitem[{Chen et~al.(2024)Chen, Chen, Liu, Steele and Zhu}]{chen3}
\bibinfo{author}{Chen, W.}, \bibinfo{author}{Chen, H.X.}, \bibinfo{author}{Liu,
  X.}, \bibinfo{author}{Steele, T.G.}, \bibinfo{author}{Zhu, S.L.},
  \bibinfo{year}{2024}.
\newblock \bibinfo{title}{Strong decays of strong decays of
  $t^a_{c{\bar{s}0}}(2900)^{++/0}$ as a fully open-flavor tetraquark state}.
\newblock \bibinfo{journal}{Eur. Phy. J. C} \bibinfo{volume}{84},
  \bibinfo{pages}{1}.
\newblock \DOIprefix\doi{10.1140/epjc/s10052-023-12355-4}.
%Type = Article
\bibitem[{Chen et~al.(2021)Chen, Han, L\"u, Wang and Yu}]{Chen:2020eyu}
\bibinfo{author}{Chen, Y.K.}, \bibinfo{author}{Han, J.J.},
  \bibinfo{author}{L\"u, Q.F.}, \bibinfo{author}{Wang, J.P.},
  \bibinfo{author}{Yu, F.S.}, \bibinfo{year}{2021}.
\newblock \bibinfo{title}{{Branching fractions of $B^-\rightarrow
  D^-X_{0,1}(2900)$ and their implications}}.
\newblock \bibinfo{journal}{Eur. Phys. J. C} \bibinfo{volume}{81},
  \bibinfo{pages}{71}.
\newblock \DOIprefix\doi{10.1140/epjc/s10052-021-08857-8},
  \href{http://arxiv.org/abs/2009.01182}{{\tt arXiv:2009.01182}}.
%Type = Article
\bibitem[{Cheng et~al.(2020)Cheng, Li, Liu, Liu, Si and Yao}]{Cheng:2020nho}
\bibinfo{author}{Cheng, J.B.}, \bibinfo{author}{Li, S.Y.},
  \bibinfo{author}{Liu, Y.R.}, \bibinfo{author}{Liu, Y.N.},
  \bibinfo{author}{Si, Z.G.}, \bibinfo{author}{Yao, T.}, \bibinfo{year}{2020}.
\newblock \bibinfo{title}{{Spectrum and rearrangement decays of tetraquark
  states with four different flavors}}.
\newblock \bibinfo{journal}{Phys. Rev. D} \bibinfo{volume}{101},
  \bibinfo{pages}{114017}.
\newblock \DOIprefix\doi{10.1103/PhysRevD.101.114017},
  \href{http://arxiv.org/abs/2001.05287}{{\tt arXiv:2001.05287}}.
%Type = Article
\bibitem[{Dong et~al.(2021a)Dong, Guo and Zou}]{Dong:2021bvy}
\bibinfo{author}{Dong, X.K.}, \bibinfo{author}{Guo, F.K.},
  \bibinfo{author}{Zou, B.S.}, \bibinfo{year}{2021}a.
\newblock \bibinfo{title}{{A survey of heavy\textendash{}heavy hadronic
  molecules}}.
\newblock \bibinfo{journal}{Commun. Theor. Phys.} \bibinfo{volume}{73},
  \bibinfo{pages}{125201}.
\newblock \DOIprefix\doi{10.1088/1572-9494/ac27a2},
  \href{http://arxiv.org/abs/2108.02673}{{\tt arXiv:2108.02673}}.
%Type = Article
\bibitem[{Dong et~al.(2021b)Dong, Guo and Zou}]{Dong:2020hxe}
\bibinfo{author}{Dong, X.K.}, \bibinfo{author}{Guo, F.K.},
  \bibinfo{author}{Zou, B.S.}, \bibinfo{year}{2021}b.
\newblock \bibinfo{title}{{Explaining the Many Threshold Structures in the
  Heavy-Quark Hadron Spectrum}}.
\newblock \bibinfo{journal}{Phys. Rev. Lett.} \bibinfo{volume}{126},
  \bibinfo{pages}{152001}.
\newblock \DOIprefix\doi{10.1103/PhysRevLett.126.152001},
  \href{http://arxiv.org/abs/2011.14517}{{\tt arXiv:2011.14517}}.
%Type = Article
\bibitem[{Dong and Zou(2021)}]{Dong:2020rgs}
\bibinfo{author}{Dong, X.K.}, \bibinfo{author}{Zou, B.S.},
  \bibinfo{year}{2021}.
\newblock \bibinfo{title}{{Prediction of possible $DK_1$ bound states}}.
\newblock \bibinfo{journal}{Eur. Phys. J. A} \bibinfo{volume}{57},
  \bibinfo{pages}{139}.
\newblock \DOIprefix\doi{10.1140/epja/s10050-021-00442-7},
  \href{http://arxiv.org/abs/2009.11619}{{\tt arXiv:2009.11619}}.
%Type = Article
\bibitem[{Garcilazo and Valcarce(2020)}]{Garcilazo:2020bgc}
\bibinfo{author}{Garcilazo, H.}, \bibinfo{author}{Valcarce, A.},
  \bibinfo{year}{2020}.
\newblock \bibinfo{title}{{Hidden and Open Heavy-Flavor Hadronic States}}.
\newblock \bibinfo{journal}{Few Body Syst.} \bibinfo{volume}{61},
  \bibinfo{pages}{24}.
\newblock \DOIprefix\doi{10.1007/s00601-020-01557-1},
  \href{http://arxiv.org/abs/2007.06046}{{\tt arXiv:2007.06046}}.
%Type = Article
\bibitem[{Gell-Mann(1964)}]{GellMann:1964nj}
\bibinfo{author}{Gell-Mann, M.}, \bibinfo{year}{1964}.
\newblock \bibinfo{title}{{A Schematic Model of Baryons and Mesons}}.
\newblock \bibinfo{journal}{Phys. Lett.} \bibinfo{volume}{8},
  \bibinfo{pages}{214--215}.
\newblock \DOIprefix\doi{10.1016/S0031-9163(64)92001-3}.
%Type = Article
\bibitem[{Gordillo and Alcaraz-Pelegrina(2023)}]{Gordillo:2023tnz}
\bibinfo{author}{Gordillo, M.C.}, \bibinfo{author}{Alcaraz-Pelegrina, J.M.},
  \bibinfo{year}{2023}.
\newblock \bibinfo{title}{{Asymptotic mass limit of large fully heavy compact
  multiquarks}}.
\newblock \bibinfo{journal}{Phys. Rev. D} \bibinfo{volume}{108},
  \bibinfo{pages}{054027}.
\newblock \DOIprefix\doi{10.1103/PhysRevD.108.054027},
  \href{http://arxiv.org/abs/2307.08408}{{\tt arXiv:2307.08408}}.
%Type = Article
\bibitem[{Gordillo et~al.(2020)Gordillo, De~Soto and
  Segovia}]{Gordillo:2020sgc}
\bibinfo{author}{Gordillo, M.C.}, \bibinfo{author}{De~Soto, F.},
  \bibinfo{author}{Segovia, J.}, \bibinfo{year}{2020}.
\newblock \bibinfo{title}{{Diffusion Monte Carlo calculations of fully-heavy
  multiquark bound states}}.
\newblock \bibinfo{journal}{Phys. Rev. D} \bibinfo{volume}{102},
  \bibinfo{pages}{114007}.
\newblock \DOIprefix\doi{10.1103/PhysRevD.102.114007},
  \href{http://arxiv.org/abs/2009.11889}{{\tt arXiv:2009.11889}}.
%Type = Article
\bibitem[{Gordillo et~al.(2021)Gordillo, De~Soto and
  Segovia}]{Gordillo:2021bra}
\bibinfo{author}{Gordillo, M.C.}, \bibinfo{author}{De~Soto, F.},
  \bibinfo{author}{Segovia, J.}, \bibinfo{year}{2021}.
\newblock \bibinfo{title}{{Structure of the X(3872) as explained by a diffusion
  Monte Carlo calculation}}.
\newblock \bibinfo{journal}{Phys. Rev. D} \bibinfo{volume}{104},
  \bibinfo{pages}{054036}.
\newblock \DOIprefix\doi{10.1103/PhysRevD.104.054036},
  \href{http://arxiv.org/abs/2105.11976}{{\tt arXiv:2105.11976}}.
%Type = Article
\bibitem[{Gordillo et~al.(2022)Gordillo, De~Soto and
  Segovia}]{Gordillo:2022nnj}
\bibinfo{author}{Gordillo, M.C.}, \bibinfo{author}{De~Soto, F.},
  \bibinfo{author}{Segovia, J.}, \bibinfo{year}{2022}.
\newblock \bibinfo{title}{{X(3872){\textquoteright}s excitation and its
  connection with production at hadron colliders}}.
\newblock \bibinfo{journal}{Phys. Rev. D} \bibinfo{volume}{106},
  \bibinfo{pages}{094004}.
\newblock \DOIprefix\doi{10.1103/PhysRevD.106.094004},
  \href{http://arxiv.org/abs/2209.04221}{{\tt arXiv:2209.04221}}.
%Type = Article
\bibitem[{Gordillo and Segovia(2024)}]{Gordillo:2024sem}
\bibinfo{author}{Gordillo, M.C.}, \bibinfo{author}{Segovia, J.},
  \bibinfo{year}{2024}.
\newblock \bibinfo{title}{{Heavy multiquark systems as clusters of smaller
  units: A diffusion Monte~Carlo calculation}}.
\newblock \bibinfo{journal}{Phys. Rev. D} \bibinfo{volume}{109},
  \bibinfo{pages}{094032}.
\newblock \DOIprefix\doi{10.1103/PhysRevD.109.094032},
  \href{http://arxiv.org/abs/2403.15000}{{\tt arXiv:2403.15000}}.
%Type = Article
\bibitem[{Gordillo et~al.(2024)Gordillo, Segovia and
  Alcaraz-Pelegrina}]{Gordillo:2024blx}
\bibinfo{author}{Gordillo, M.C.}, \bibinfo{author}{Segovia, J.},
  \bibinfo{author}{Alcaraz-Pelegrina, J.M.}, \bibinfo{year}{2024}.
\newblock \bibinfo{title}{{Diffusion Monte~Carlo calculation of fully heavy
  pentaquarks}}.
\newblock \bibinfo{journal}{Phys. Rev. D} \bibinfo{volume}{110},
  \bibinfo{pages}{094024}.
\newblock \DOIprefix\doi{10.1103/PhysRevD.110.094024},
  \href{http://arxiv.org/abs/2409.04130}{{\tt arXiv:2409.04130}}.
%Type = Article
\bibitem[{Guo et~al.(2018)Guo, Hanhart, Mei\ss{}ner, Wang, Zhao and
  Zou}]{Guo:2017jvc}
\bibinfo{author}{Guo, F.K.}, \bibinfo{author}{Hanhart, C.},
  \bibinfo{author}{Mei\ss{}ner, U.G.}, \bibinfo{author}{Wang, Q.},
  \bibinfo{author}{Zhao, Q.}, \bibinfo{author}{Zou, B.S.},
  \bibinfo{year}{2018}.
\newblock \bibinfo{title}{{Hadronic molecules}}.
\newblock \bibinfo{journal}{Rev. Mod. Phys.} \bibinfo{volume}{90},
  \bibinfo{pages}{015004}.
\newblock \DOIprefix\doi{10.1103/RevModPhys.90.015004},
  \href{http://arxiv.org/abs/1705.00141}{{\tt arXiv:1705.00141}}.
%Type = Book
\bibitem[{Hammond et~al.(1994)Hammond, Lester and Reynolds}]{Hammond:1994bk}
\bibinfo{author}{Hammond, B.}, \bibinfo{author}{Lester, W.},
  \bibinfo{author}{Reynolds, P.}, \bibinfo{year}{1994}.
\newblock \bibinfo{title}{Monte Carlo Methods in ab Initio Quantum Chemistry}.
\newblock \bibinfo{publisher}{World Scientific}, \bibinfo{address}{Singapore}.
%Type = Article
\bibitem[{He and Chen(2021)}]{He:2020btl}
\bibinfo{author}{He, J.}, \bibinfo{author}{Chen, D.Y.}, \bibinfo{year}{2021}.
\newblock \bibinfo{title}{{Molecular picture for $X_0(2900)$ and $X_1(2900)$}}.
\newblock \bibinfo{journal}{Chin. Phys. C} \bibinfo{volume}{45},
  \bibinfo{pages}{063102}.
\newblock \DOIprefix\doi{10.1088/1674-1137/abeda8},
  \href{http://arxiv.org/abs/2008.07782}{{\tt arXiv:2008.07782}}.
%Type = Article
\bibitem[{He et~al.(2020)He, Wang and Zhu}]{He:2020jna}
\bibinfo{author}{He, X.G.}, \bibinfo{author}{Wang, W.}, \bibinfo{author}{Zhu,
  R.}, \bibinfo{year}{2020}.
\newblock \bibinfo{title}{{Open-charm tetraquark $X_c$ and open-bottom
  tetraquark $X_b$}}.
\newblock \bibinfo{journal}{Eur. Phys. J. C} \bibinfo{volume}{80},
  \bibinfo{pages}{1026}.
\newblock \DOIprefix\doi{10.1140/epjc/s10052-020-08597-1},
  \href{http://arxiv.org/abs/2008.07145}{{\tt arXiv:2008.07145}}.
%Type = Article
\bibitem[{Hu et~al.(2021)Hu, Lao, Ling and Wang}]{Hu:2020mxp}
\bibinfo{author}{Hu, M.W.}, \bibinfo{author}{Lao, X.Y.}, \bibinfo{author}{Ling,
  P.}, \bibinfo{author}{Wang, Q.}, \bibinfo{year}{2021}.
\newblock \bibinfo{title}{{$X_0$(2900) and its heavy quark spin partners in
  molecular picture}}.
\newblock \bibinfo{journal}{Chin. Phys. C} \bibinfo{volume}{45},
  \bibinfo{pages}{021003}.
\newblock \DOIprefix\doi{10.1088/1674-1137/abcfaa},
  \href{http://arxiv.org/abs/2008.06894}{{\tt arXiv:2008.06894}}.
%Type = Article
\bibitem[{J.M.~Richard(2017)}]{vijande}
\bibinfo{author}{J.M.~Richard, A.Valcarce, J.}, \bibinfo{year}{2017}.
\newblock \bibinfo{title}{{Stable heavy pentaquarks in constituent models}}.
\newblock \bibinfo{journal}{Phys. Lett. B} \bibinfo{volume}{774},
  \bibinfo{pages}{710--714}.
\newblock \DOIprefix\doi{10.1016/j.physletb.2017.10.036},
  \href{http://arxiv.org/abs/1710.08239}{{\tt arXiv:1710.08239}}.
%Type = Article
\bibitem[{Karliner and Rosner(2020)}]{Karliner:2020vsi}
\bibinfo{author}{Karliner, M.}, \bibinfo{author}{Rosner, J.L.},
  \bibinfo{year}{2020}.
\newblock \bibinfo{title}{{First exotic hadron with open heavy flavor: $cs\bar
  u\bar d$ tetraquark}}.
\newblock \bibinfo{journal}{Phys. Rev. D} \bibinfo{volume}{102},
  \bibinfo{pages}{094016}.
\newblock \DOIprefix\doi{10.1103/PhysRevD.102.094016},
  \href{http://arxiv.org/abs/2008.05993}{{\tt arXiv:2008.05993}}.
%Type = Article
\bibitem[{Liu et~al.(2020a)Liu, Xie and Geng}]{Liu:2020nil}
\bibinfo{author}{Liu, M.Z.}, \bibinfo{author}{Xie, J.J.},
  \bibinfo{author}{Geng, L.S.}, \bibinfo{year}{2020}a.
\newblock \bibinfo{title}{{$X_0(2866)$ as a $D^*\bar{K}^*$ molecular state}}.
\newblock \bibinfo{journal}{Phys. Rev. D} \bibinfo{volume}{102},
  \bibinfo{pages}{091502}.
\newblock \DOIprefix\doi{10.1103/PhysRevD.102.091502},
  \href{http://arxiv.org/abs/2008.07389}{{\tt arXiv:2008.07389}}.
%Type = Article
\bibitem[{Liu et~al.(2020b)Liu, Yan, Ke, Li and Xie}]{Liu:2020orv}
\bibinfo{author}{Liu, X.H.}, \bibinfo{author}{Yan, M.J.}, \bibinfo{author}{Ke,
  H.W.}, \bibinfo{author}{Li, G.}, \bibinfo{author}{Xie, J.J.},
  \bibinfo{year}{2020}b.
\newblock \bibinfo{title}{{Triangle singularity as the origin of $X_0(2900)$
  and $X_1(2900)$ observed in $B^+\to D^+ D^- K^+$}}.
\newblock \bibinfo{journal}{Eur. Phys. J. C} \bibinfo{volume}{80},
  \bibinfo{pages}{1178}.
\newblock \DOIprefix\doi{10.1140/epjc/s10052-020-08762-6},
  \href{http://arxiv.org/abs/2008.07190}{{\tt arXiv:2008.07190}}.
%Type = Article
\bibitem[{Liu et~al.(2019)Liu, Chen, Chen, Liu and Zhu}]{Liu:2019zoy}
\bibinfo{author}{Liu, Y.R.}, \bibinfo{author}{Chen, H.X.},
  \bibinfo{author}{Chen, W.}, \bibinfo{author}{Liu, X.}, \bibinfo{author}{Zhu,
  S.L.}, \bibinfo{year}{2019}.
\newblock \bibinfo{title}{{Pentaquark and Tetraquark states}}.
\newblock \bibinfo{journal}{Prog. Part. Nucl. Phys.} \bibinfo{volume}{107},
  \bibinfo{pages}{237--320}.
\newblock \DOIprefix\doi{10.1016/j.ppnp.2019.04.003},
  \href{http://arxiv.org/abs/1903.11976}{{\tt arXiv:1903.11976}}.
%Type = Article
\bibitem[{Ma et~al.(2023a)Ma, Meng, Chen and Zhu}]{baryons}
\bibinfo{author}{Ma, Y.}, \bibinfo{author}{Meng, L.}, \bibinfo{author}{Chen,
  Y.K.}, \bibinfo{author}{Zhu, S.L.}, \bibinfo{year}{2023}a.
\newblock \bibinfo{title}{Ground state baryons in the flux-tube three-body
  confinement model using diffusion monte carlo}.
\newblock \bibinfo{journal}{Phys. Rev. D} \bibinfo{volume}{107},
  \bibinfo{pages}{054035}.
\newblock \URLprefix
  \url{https://link.aps.org/doi/10.1103/PhysRevD.107.054035},
  \DOIprefix\doi{10.1103/PhysRevD.107.054035}.
%Type = Article
\bibitem[{Ma et~al.(2023b)Ma, Meng, Chen and Zhu}]{Ma:2022vqf}
\bibinfo{author}{Ma, Y.}, \bibinfo{author}{Meng, L.}, \bibinfo{author}{Chen,
  Y.K.}, \bibinfo{author}{Zhu, S.L.}, \bibinfo{year}{2023}b.
\newblock \bibinfo{title}{{Ground state baryons in the flux-tube three-body
  confinement model using diffusion Monte~Carlo}}.
\newblock \bibinfo{journal}{Phys. Rev. D} \bibinfo{volume}{107},
  \bibinfo{pages}{054035}.
\newblock \DOIprefix\doi{10.1103/PhysRevD.107.054035},
  \href{http://arxiv.org/abs/2211.09021}{{\tt arXiv:2211.09021}}.
%Type = Article
\bibitem[{Ma et~al.(2024)Ma, Meng, Chen and Zhu}]{Ma:2023int}
\bibinfo{author}{Ma, Y.}, \bibinfo{author}{Meng, L.}, \bibinfo{author}{Chen,
  Y.K.}, \bibinfo{author}{Zhu, S.L.}, \bibinfo{year}{2024}.
\newblock \bibinfo{title}{{Doubly heavy tetraquark states in the constituent
  quark model using diffusion Monte~Carlo method}}.
\newblock \bibinfo{journal}{Phys. Rev. D} \bibinfo{volume}{109},
  \bibinfo{pages}{074001}.
\newblock \DOIprefix\doi{10.1103/PhysRevD.109.074001},
  \href{http://arxiv.org/abs/2309.17068}{{\tt arXiv:2309.17068}}.
%Type = Article
\bibitem[{Mai et~al.(2023)Mai, Mei\ss{}ner and Urbach}]{Mai:2022eur}
\bibinfo{author}{Mai, M.}, \bibinfo{author}{Mei\ss{}ner, U.G.},
  \bibinfo{author}{Urbach, C.}, \bibinfo{year}{2023}.
\newblock \bibinfo{title}{{Towards a theory of hadron resonances}}.
\newblock \bibinfo{journal}{Phys. Rept.} \bibinfo{volume}{1001},
  \bibinfo{pages}{1--66}.
\newblock \DOIprefix\doi{10.1016/j.physrep.2022.11.005},
  \href{http://arxiv.org/abs/2206.01477}{{\tt arXiv:2206.01477}}.
%Type = Article
\bibitem[{Meng et~al.(2023)Meng, Chen, Ma and Zhu}]{mesons}
\bibinfo{author}{Meng, L.}, \bibinfo{author}{Chen, Y.K.}, \bibinfo{author}{Ma,
  Y.}, \bibinfo{author}{Zhu, S.L.}, \bibinfo{year}{2023}.
\newblock \bibinfo{title}{Tetraquark bound states in constituent quark models:
  Benchmark test calculations}.
\newblock \bibinfo{journal}{Phys. Rev. D} \bibinfo{volume}{108},
  \bibinfo{pages}{114016}.
\newblock \URLprefix
  \url{https://link.aps.org/doi/10.1103/PhysRevD.108.114016},
  \DOIprefix\doi{10.1103/PhysRevD.108.114016}.
%Type = Article
\bibitem[{Molina et~al.(2010)Molina, Branz and Oset}]{Molina:2010tx}
\bibinfo{author}{Molina, R.}, \bibinfo{author}{Branz, T.},
  \bibinfo{author}{Oset, E.}, \bibinfo{year}{2010}.
\newblock \bibinfo{title}{{A new interpretation for the $D^*_{s2}(2573)$ and
  the prediction of novel exotic charmed mesons}}.
\newblock \bibinfo{journal}{Phys. Rev. D} \bibinfo{volume}{82},
  \bibinfo{pages}{014010}.
\newblock \DOIprefix\doi{10.1103/PhysRevD.82.014010},
  \href{http://arxiv.org/abs/1005.0335}{{\tt arXiv:1005.0335}}.
%Type = Article
\bibitem[{Molina and Oset(2020)}]{Molina:2020hde}
\bibinfo{author}{Molina, R.}, \bibinfo{author}{Oset, E.}, \bibinfo{year}{2020}.
\newblock \bibinfo{title}{{Molecular picture for the $X_0(2866)$ as a $D^*
  \bar{K}^*$ $J^P=0^+$ state and related $1^+,2^+$ states}}.
\newblock \bibinfo{journal}{Phys. Lett. B} \bibinfo{volume}{811},
  \bibinfo{pages}{135870}.
\newblock \DOIprefix\doi{10.1016/j.physletb.2020.135870},
  \href{http://arxiv.org/abs/2008.11171}{{\tt arXiv:2008.11171}}.
%Type = Article
\bibitem[{Molina and Oset(2023)}]{Molina:2022jcd}
\bibinfo{author}{Molina, R.}, \bibinfo{author}{Oset, E.}, \bibinfo{year}{2023}.
\newblock \bibinfo{title}{{Tcs\textasciimacron{}(2900) as a threshold effect
  from the interaction of the D*K*, Ds*\ensuremath{\rho} channels}}.
\newblock \bibinfo{journal}{Phys. Rev. D} \bibinfo{volume}{107},
  \bibinfo{pages}{056015}.
\newblock \DOIprefix\doi{10.1103/PhysRevD.107.056015},
  \href{http://arxiv.org/abs/2211.01302}{{\tt arXiv:2211.01302}}.
%Type = Article
\bibitem[{Mutuk(2024)}]{Mutuk:2023oyz}
\bibinfo{author}{Mutuk, H.}, \bibinfo{year}{2024}.
\newblock \bibinfo{title}{{Masses and magnetic moments of doubly heavy
  tetraquarks via diffusion Monte Carlo method}}.
\newblock \bibinfo{journal}{Eur. Phys. J. C} \bibinfo{volume}{84},
  \bibinfo{pages}{395}.
\newblock \DOIprefix\doi{10.1140/epjc/s10052-024-12736-3},
  \href{http://arxiv.org/abs/2312.13383}{{\tt arXiv:2312.13383}}.
%Type = Article
\bibitem[{Navas(2024)}]{pdg2024}
\bibinfo{author}{Navas, S.e.a.} (\bibinfo{collaboration}{Particle Data Group
  Collaboration}), \bibinfo{year}{2024}.
\newblock \bibinfo{title}{Review of particle physics}.
\newblock \bibinfo{journal}{Phys. Rev. D} \bibinfo{volume}{110},
  \bibinfo{pages}{030001}.
\newblock \URLprefix
  \url{https://link.aps.org/doi/10.1103/PhysRevD.110.030001},
  \DOIprefix\doi{10.1103/PhysRevD.110.030001}.
%Type = Article
\bibitem[{Ortega and Entem(2021)}]{Ortega:2020tng}
\bibinfo{author}{Ortega, P.G.}, \bibinfo{author}{Entem, D.R.},
  \bibinfo{year}{2021}.
\newblock \bibinfo{title}{{Coupling hadron-hadron thresholds within a chiral
  quark model approach}}.
\newblock \bibinfo{journal}{Symmetry} \bibinfo{volume}{13},
  \bibinfo{pages}{279}.
\newblock \DOIprefix\doi{10.3390/sym13020279},
  \href{http://arxiv.org/abs/2012.10105}{{\tt arXiv:2012.10105}}.
%Type = Article
\bibitem[{Ortega et~al.(2023)Ortega, Entem, Fernandez and
  Segovia}]{Ortega:2023azl}
\bibinfo{author}{Ortega, P.G.}, \bibinfo{author}{Entem, D.R.},
  \bibinfo{author}{Fernandez, F.}, \bibinfo{author}{Segovia, J.},
  \bibinfo{year}{2023}.
\newblock \bibinfo{title}{{Novel Tcs and Tcs\textasciimacron{} candidates in a
  constituent-quark-model-based meson-meson coupled-channels calculation}}.
\newblock \bibinfo{journal}{Phys. Rev. D} \bibinfo{volume}{108},
  \bibinfo{pages}{094035}.
\newblock \DOIprefix\doi{10.1103/PhysRevD.108.094035},
  \href{http://arxiv.org/abs/2305.14430}{{\tt arXiv:2305.14430}}.
%Type = Article
\bibitem[{Segovia et~al.(2011)Segovia, Albertus, Entem, Fern\'andez,
  Hern\'andez and P\'erez-Garc\'{\i}a}]{jorgeSLM}
\bibinfo{author}{Segovia, J.}, \bibinfo{author}{Albertus, C.},
  \bibinfo{author}{Entem, D.R.}, \bibinfo{author}{Fern\'andez, F.},
  \bibinfo{author}{Hern\'andez, E.}, \bibinfo{author}{P\'erez-Garc\'{\i}a,
  M.A.}, \bibinfo{year}{2011}.
\newblock \bibinfo{title}{Semileptonic $b$ and ${B}_{s}$ decays into orbitally
  excited charmed mesons}.
\newblock \bibinfo{journal}{Phys. Rev. D} \bibinfo{volume}{84},
  \bibinfo{pages}{094029}.
\newblock \URLprefix \url{https://link.aps.org/doi/10.1103/PhysRevD.84.094029},
  \DOIprefix\doi{10.1103/PhysRevD.84.094029}.
%Type = Article
\bibitem[{Semay and Silvestre-Brac(1994)}]{Semay:1994ht}
\bibinfo{author}{Semay, C.}, \bibinfo{author}{Silvestre-Brac, B.},
  \bibinfo{year}{1994}.
\newblock \bibinfo{title}{{Diquonia and potential models}}.
\newblock \bibinfo{journal}{Z. Phys. C} \bibinfo{volume}{61},
  \bibinfo{pages}{271--275}.
\newblock \DOIprefix\doi{10.1007/BF01413104}.
%Type = Article
\bibitem[{Silvestre-Brac(1996)}]{Silvestre-Brac:1996myf}
\bibinfo{author}{Silvestre-Brac, B.}, \bibinfo{year}{1996}.
\newblock \bibinfo{title}{{Spectrum and static properties of heavy baryons}}.
\newblock \bibinfo{journal}{Few Body Syst.} \bibinfo{volume}{20},
  \bibinfo{pages}{1--25}.
\newblock \DOIprefix\doi{10.1007/s006010050028}.
%Type = Article
\bibitem[{Tan and Ping(2021)}]{Tan:2020cpu}
\bibinfo{author}{Tan, Y.}, \bibinfo{author}{Ping, J.}, \bibinfo{year}{2021}.
\newblock \bibinfo{title}{{X(2900) in a chiral quark model}}.
\newblock \bibinfo{journal}{Chin. Phys. C} \bibinfo{volume}{45},
  \bibinfo{pages}{093104}.
\newblock \DOIprefix\doi{10.1088/1674-1137/ac0ba4},
  \href{http://arxiv.org/abs/2010.04045}{{\tt arXiv:2010.04045}}.
%Type = Article
\bibitem[{Wang et~al.(2021)Wang, Meng, Xiao, Oka and Zhu}]{Wang:2020prk}
\bibinfo{author}{Wang, G.J.}, \bibinfo{author}{Meng, L.},
  \bibinfo{author}{Xiao, L.Y.}, \bibinfo{author}{Oka, M.},
  \bibinfo{author}{Zhu, S.L.}, \bibinfo{year}{2021}.
\newblock \bibinfo{title}{{Mass spectrum and strong decays of tetraquark
  ${\bar{c}}{\bar{s}} qq$ states}}.
\newblock \bibinfo{journal}{Eur. Phys. J. C} \bibinfo{volume}{81},
  \bibinfo{pages}{188}.
\newblock \DOIprefix\doi{10.1140/epjc/s10052-021-08978-0},
  \href{http://arxiv.org/abs/2010.09395}{{\tt arXiv:2010.09395}}.
%Type = Article
\bibitem[{Wang(2020)}]{Wang:2020xyc}
\bibinfo{author}{Wang, Z.G.}, \bibinfo{year}{2020}.
\newblock \bibinfo{title}{{Analysis of the $X_0(2900)$ as the scalar tetraquark
  state via the QCD sum rules}}.
\newblock \bibinfo{journal}{Int. J. Mod. Phys. A} \bibinfo{volume}{35},
  \bibinfo{pages}{2050187}.
\newblock \DOIprefix\doi{10.1142/S0217751X20501870},
  \href{http://arxiv.org/abs/2008.07833}{{\tt arXiv:2008.07833}}.
%Type = Article
\bibitem[{Wei et~al.(2022)Wei, Wang, An and Deng}]{Wei:2022wtr}
\bibinfo{author}{Wei, J.}, \bibinfo{author}{Wang, Y.H.}, \bibinfo{author}{An,
  C.S.}, \bibinfo{author}{Deng, C.R.}, \bibinfo{year}{2022}.
\newblock \bibinfo{title}{{Color flux-tube nature of the states Tcs(2900) and
  Tcs\textasciimacron{}a(2900)}}.
\newblock \bibinfo{journal}{Phys. Rev. D} \bibinfo{volume}{106},
  \bibinfo{pages}{096023}.
\newblock \DOIprefix\doi{10.1103/PhysRevD.106.096023},
  \href{http://arxiv.org/abs/2210.04841}{{\tt arXiv:2210.04841}}.
%Type = Article
\bibitem[{Xiao et~al.(2021)Xiao, Chen, Dong and Meng}]{Xiao:2020ltm}
\bibinfo{author}{Xiao, C.J.}, \bibinfo{author}{Chen, D.Y.},
  \bibinfo{author}{Dong, Y.B.}, \bibinfo{author}{Meng, G.W.},
  \bibinfo{year}{2021}.
\newblock \bibinfo{title}{{Study of the decays of $S-$wave $\bar D^\ast K^\ast$
  hadronic molecules: The scalar $X_0(2900)$ and its spin partners
  $X_{J(J=1,2)}$}}.
\newblock \bibinfo{journal}{Phys. Rev. D} \bibinfo{volume}{103},
  \bibinfo{pages}{034004}.
\newblock \DOIprefix\doi{10.1103/PhysRevD.103.034004},
  \href{http://arxiv.org/abs/2009.14538}{{\tt arXiv:2009.14538}}.
%Type = Article
\bibitem[{Xue et~al.(2021)Xue, Jin, Huang and Ping}]{Xue:2020vtq}
\bibinfo{author}{Xue, Y.}, \bibinfo{author}{Jin, X.}, \bibinfo{author}{Huang,
  H.}, \bibinfo{author}{Ping, J.}, \bibinfo{year}{2021}.
\newblock \bibinfo{title}{{Tetraquarks with open charm flavor}}.
\newblock \bibinfo{journal}{Phys. Rev. D} \bibinfo{volume}{103},
  \bibinfo{pages}{054010}.
\newblock \DOIprefix\doi{10.1103/PhysRevD.103.054010},
  \href{http://arxiv.org/abs/2008.09516}{{\tt arXiv:2008.09516}}.
%Type = Article
\bibitem[{Yang et~al.(2020)Yang, Ping and Segovia}]{Yang:2020atz}
\bibinfo{author}{Yang, G.}, \bibinfo{author}{Ping, J.},
  \bibinfo{author}{Segovia, J.}, \bibinfo{year}{2020}.
\newblock \bibinfo{title}{{Tetra- and penta-quark structures in the constituent
  quark model}}.
\newblock \bibinfo{journal}{Symmetry} \bibinfo{volume}{12},
  \bibinfo{pages}{1869}.
\newblock \DOIprefix\doi{10.3390/sym12111869},
  \href{http://arxiv.org/abs/2009.00238}{{\tt arXiv:2009.00238}}.
%Type = Article
\bibitem[{Yang et~al.(2021)Yang, Ping and Segovia}]{Yang:2021izl}
\bibinfo{author}{Yang, G.}, \bibinfo{author}{Ping, J.},
  \bibinfo{author}{Segovia, J.}, \bibinfo{year}{2021}.
\newblock \bibinfo{title}{{$\mathbf{sQ\bar{q}\bar{q}}$ $\mathbf{(q=u,\,d;\,
  Q=c,\,b)}$ tetraquarks in the chiral quark model}}.
\newblock \bibinfo{journal}{Phys. Rev. D} \bibinfo{volume}{103},
  \bibinfo{pages}{074011}.
\newblock \DOIprefix\doi{10.1103/PhysRevD.103.074011},
  \href{http://arxiv.org/abs/2101.04933}{{\tt arXiv:2101.04933}}.
%Type = Article
\bibitem[{Yang et~al.(2023)Yang, Xin and Wang}]{Yang:2023evp}
\bibinfo{author}{Yang, X.S.}, \bibinfo{author}{Xin, Q.}, \bibinfo{author}{Wang,
  Z.G.}, \bibinfo{year}{2023}.
\newblock \bibinfo{title}{{Analysis of the Tcs(2900) and related tetraquark
  states with the QCD sum rules}}.
\newblock \bibinfo{journal}{Int. J. Mod. Phys. A} \bibinfo{volume}{38},
  \bibinfo{pages}{2350056}.
\newblock \DOIprefix\doi{10.1142/S0217751X23500562},
  \href{http://arxiv.org/abs/2302.01718}{{\tt arXiv:2302.01718}}.
%Type = Article
\bibitem[{Yue et~al.(2023)Yue, Xiao and Chen}]{Yue:2022mnf}
\bibinfo{author}{Yue, Z.L.}, \bibinfo{author}{Xiao, C.J.},
  \bibinfo{author}{Chen, D.Y.}, \bibinfo{year}{2023}.
\newblock \bibinfo{title}{{Decays of the fully open flavor state
  Tcs\textasciimacron{}00 in a D*K* molecule scenario}}.
\newblock \bibinfo{journal}{Phys. Rev. D} \bibinfo{volume}{107},
  \bibinfo{pages}{034018}.
\newblock \DOIprefix\doi{10.1103/PhysRevD.107.034018},
  \href{http://arxiv.org/abs/2212.03018}{{\tt arXiv:2212.03018}}.
%Type = Article
\bibitem[{Zhang(2021)}]{Zhang:2020oze}
\bibinfo{author}{Zhang, J.R.}, \bibinfo{year}{2021}.
\newblock \bibinfo{title}{{Open-charm tetraquark candidate: Note on
  $X_0$(2900)}}.
\newblock \bibinfo{journal}{Phys. Rev. D} \bibinfo{volume}{103},
  \bibinfo{pages}{054019}.
\newblock \DOIprefix\doi{10.1103/PhysRevD.103.054019},
  \href{http://arxiv.org/abs/2008.07295}{{\tt arXiv:2008.07295}}.
%Type = Article
\bibitem[{Zweig(1964)}]{Zweig:1964CERN}
\bibinfo{author}{Zweig, G.}, \bibinfo{year}{1964}.
\newblock \bibinfo{title}{Developments in the quark theory of hadrons}.
\newblock \bibinfo{journal}{CERN Report No.8182/TH.401, CERN Report
  No.8419/TH.412} .

\end{thebibliography}



%% For example:

%% \bibitem[Aladro et al.(2015)]{Aladro15} Aladro, R., Martín, S., Riquelme, D., et al. 2015, \aas, 579, A101

%%\end{thebibliography}

\end{document}